\documentclass[epsf]{elsart}

 \usepackage{epsfig}

\usepackage{amssymb}

\begin{document}

\begin{frontmatter}



\title{Development of Atmospheric Monitoring System at Akeno Observatory 
	for the Telescope Array Project}


\author[1]{T.Yamamoto}
\author[2]{M.Chikawa}
\author[1]{N.Hayashida}
\author[4]{S.Kawakami}
\author[1]{N.Minagawa}
\author[2]{Y.Morizane}
\author[3]{M.Sasano}
\author[1]{M.Teshima}
\author[2]{K.Yasui}
\author[]{and The Telescope Array collaboration}

\address[1]{Institute for Cosmic Ray Research, University of Tokyo,Tokyo, Japan}
\address[2]{Department of Physics, Kinki University, Higashi-Osaka, Japan}
\address[3]{Communication Research Laboratory, Tokyo, Japan}
\address[4]{Department of Physics, Osaka City University, Osaka, Japan}

\begin{abstract}

 We have developed an atmospheric monitoring system for the Telescope Array
 experiment at Akeno Observatory. It consists of a Nd:YAG laser with
 an alt-azimuth shooting system and a small light receiver. This system is
 installed inside an air conditioned weather-proof dome. All parts, including 
 the dome, laser, shooter, receiver, and optical devices are fully
 controlled by a personal computer utilizing the Linux operating system. 
 It is now
 operated as a back-scattering LIDAR System. For the Telescope Array
 experiment, to estimate energy reliably and to obtain the correct shower
 development profile, the light
 transmittance in the atmosphere needs to be calibrated with high accuracy. 
 Based on 
 observational results using this monitoring system, we consider this
 LIDAR to be a very powerful technique for Telescope Array
 experiments. The details of this system and its atmospheric monitoring
 technique will be discussed.

\end{abstract}

\begin{keyword}
Cosmic Ray \sep Fluorescence Technique \sep Atmospheric Monitoring


\end{keyword}

\end{frontmatter}

\section{Introduction}
\label{Introduction}

%
%
\begin{figure}[htbp]
\centering
\leavevmode
\epsfxsize=14.5cm
\epsfbox{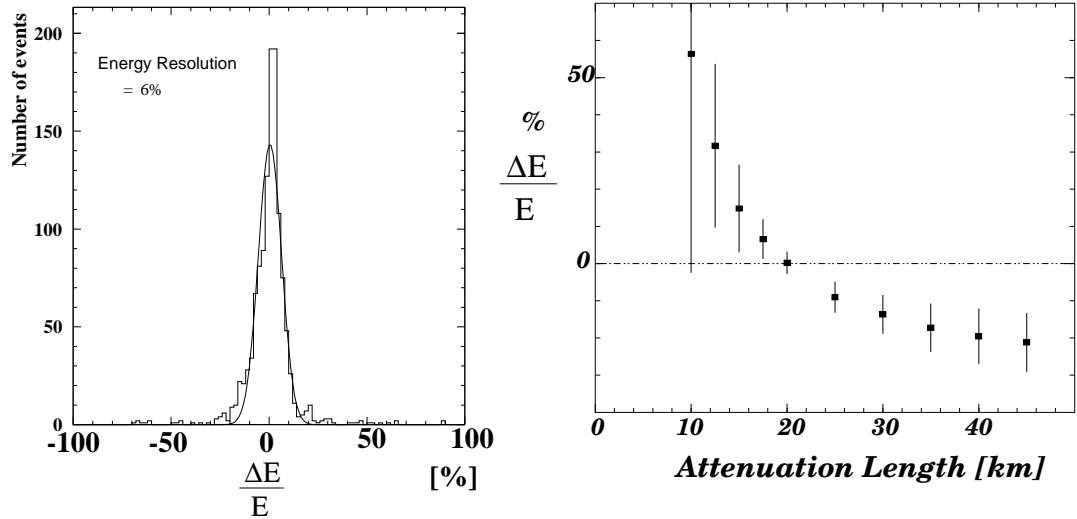}
\caption{Results of Monte-Carlo simulation for the Telescope Array. Left
 panel indicates the energy resolution of the Telescope Array for $10^{20}$
 eV cosmic rays. The energy resolution is better than 6\% in 1 $\sigma$
 as long as we know the transmittance of the atmosphere. Right panel shows
 the error of estimated energy for $10^{20}$ eV cosmic rays as a function of 
 incorrectness of atmospheric correction of attenuation length. 
 The values used for the simulation were $L_M=20$km and $H_M=1.2$km.
 }
\label{fig:ta_sim}
\end{figure}
%
%

The air fluorescence technique for air shower observation has many
advantages, (for example, direct measurement of shower longitudinal
development, and stereo geometrical reconstruction with multiple eyes). These 
advantages will bring technical breakthroughs in EHE Cosmic Ray
physics. 

On the other hand, this fluorescence technique has a serious
problem. In this technique the light
yielded from air showers after the transmission is measured in atmosphere at a
10 $\sim$ 60 km distance. Atmosphere can be considered a part of
the detector. Without any monitoring, the atmospheric conditions 
may cause large uncertainty in the experiment and result in
serious systematic error.
Therefore, the calibration of the light transmittance in the
atmosphere is essential for the air fluorescence experiment. 

Charged particles in the air shower excite air molecules 
and yield fluorescence light of between 330nm and 400nm. 
This fluorescence light is also scattered by molecules and aerosols
in the air before it reaches the detector.  
The scattering process caused by molecule is called Rayleigh scattering.
The attenuation length of this scattering process 
$X_r$ is 2974 $g/cm^2$ for a 400nm wave length. Amount of
scattered photons with wave length of $\lambda$ can thus be calculated by
following equation
\begin{equation}
\frac{dN_r}{dl} = -\rho \frac{N_r}{X_r}(\frac{400nm}{\lambda})^4
\end{equation}
where $\rho$ is the atmospheric density. In this equation, the number of
scattered photons is proportional to $\lambda^{-4}$. 
This scattering process seriously effects the observation 
of fluorescence light in the air.

For example, when we consider the typical measurement of 
fluorescence light from an air shower at
a distance of 30 km from detector, with a
detector altitude of 1.5km and an elevation angle of
30 degrees, 
the fluorescence light has to pass
through about 2200 $g/cm^2$ in the air and 70\% of the photons will 
be scattered by this Rayleigh process.

There is another scattering process, Mie scattering.
There are many kinds of scattering materials in the atmosphere, such as, 
water vapors, mists, clouds, dusts, blown small sands, 
artificial fumes, exhaust gas from cars, and smoke from forest fires. 
Furthermore, 'Kosa' from the Gobi desert in China is spread over a wide area 
of the northern hemisphere. These atmospheric aerosol's densities and 
their components change significantly 
with location and time. The scattering of photons by 
these aerosols is called Mie Scattering. In the Utah desert, 
the typical scale height of Mie scattering $H_M$ is 1.2 km and 
horizontal attenuation length $L_M$ is 20 km. 
The amount of scattering photons can be expressed by the following
equation
\begin{equation}
\frac{dN_r}{dl} = -\frac{N_r}{L_M} \times exp(-\frac{h}{H_M})
\end{equation}
Therefore, under the previously described conditions, approximately
27\% of the photons contained in the fluorescence light from the
air shower are scattered by the Mie process.
When we take into account Rayleigh and Mie scattering, 
about 22\% of the photons can reach our detector.

The Telescope Array (TA) will observe the air showers which flash
far from the detector, for example, at more than 50km.
Figure \ref{fig:ta_sim} shows the
results of Monte-Carlo simulation for TA.  According to this simulation
study, the energy resolution of TA is better than 6\% if
atmospheric transmittance is accurately taken into account. 
However, if we apply
the 20\% shifted value in the attenuation length, the estimated
cosmic ray energy has a systematic error of nearly 10\%. 
To estimate the primary composition, TA will use
$X_{max}$ which is the maximum position in the longitudinal
shower development. The intrinsic resolution of $X_{max}$ is estimated
to be 20g in the TA experiment using Monte Carlo simulation. 
If the assumed attenuation length differs by 20\%,
we estimate $X_{max}$ with a 10g systematic error. 
The correction of atmospheric transmittance affects not only energy 
determination but also air shower reconstruction. 
To realize this high resolution power of TA, 
we need to measure the attenuation length of the atmosphere 
with an accuracy of a few percent or better.

For the purpose of the atmospheric monitoring, TA and High Resolution
Fly's Eyes are  developing various methods which use a
laser or flashers\cite{law99}\cite{og1}\cite{og2}\cite{og3}\cite{og4}. 
One of well-known technique for atmospheric monitoring is LIDAR
which has been developed for environmental science. 
Using this technique, atmospheric conditions can be
monitored remotely by measuring back-scattered light from a pulsed
laser beam.

We are developing a steerable LIDAR system to study the atmospheric
monitoring method in Akeno Observatory. In the present paper, we will 
describe the details of this system and discuss how to accurately measure 
transmittance in the atmosphere.

\section{LIDAR equation}
\label{ss:lidar_eq}

In general, the intensity of the back-scattered light detected by
LIDAR is expressed by the following equation (LIDAR equation)
\begin{equation}
P(R) = \frac{P_0  \kappa  A  (c  \tau/2)  \beta(R) 
         T^{2}(R)  Y(R)}{R^2}
         + P_b
\end{equation}
where \\
$P(R)$ : intensity of the detected light  \\
$P_0$ : laser intensity  \\
$P_b$ : intensity of back ground photons  \\
$c$  : light velocity  \\
$R$ : distance from laser to target  \\
$\tau$  : integration time  \\
$Y(R)$ : geometrical efficiency of the beam track and receiver  \\
$A$  : aperture of receiver  \\
$\kappa$  : detector efficiency  \\
\underline{$\beta$  : back-scatter coefficient}  \\
\underline{$T(R)$  : transmittance (=$exp(-\int^{R}_{0}\alpha dr)$)}  \\
$\alpha$  : extinction coefficient (=1/attenuation length)  \\ \\
We define the following parameters for convenience
\begin{equation}
P_0 \kappa A \frac{c\tau}{2} Y(R) \equiv C
\end{equation}
\begin{equation}
X(R) \equiv R^2(P(R) - P_b) \\
\end{equation}
\begin{equation}
S(R)\equiv ln(X(R))
\end{equation}
The LIDAR equation can be then written as follows:
\begin{equation}
X(R) = C \beta(R)T^2(R) = C \beta(R)\cdot exp(-2\int^{R}_{0} \alpha dr) 
\end{equation}
If $\alpha$ is constant until $R$,
\begin{equation}
X(R)=C\beta exp(-2R\alpha)
\end{equation}
Then we obtain the differential LIDAR equation
\begin{equation}
\frac{dS(R)}{dR} = \frac{1}{\beta(R)} \frac{d\beta(R)}{dR} - 2\alpha(R)
\label{eq:lidar}
\end{equation}
There are two variable parameters in this LIDAR equation. One is
the extinction coefficient $\alpha$ that represents the amount of scattered
photons in a scattering volume. This value corresponds to the reciprocal of 
attenuation length. The other variable parameter is the back-scattering
coefficient $\beta$ that represents the amount of scattered photons in the
backward direction. 
This value is proportional to the density of scattering matter and
its differential cross-section. Since the LIDAR equation has two
variables, it can not be solved in general.

If the atmosphere is uniform, $\beta$ is constant and $d\beta/dR=0$. Then
\begin{equation}
\label{eq:slobe}
\alpha=-\frac{1}{2}\frac{dS}{dR}
\end{equation}
Thus we can measure the $\alpha$ using this equation.

If we assume a relationship between $\alpha$ and $\beta$, the variable
parameters can be reduced. For this purpose, an empirical relationship is
proposed as follows:
\begin{equation}
\beta = const\cdot \alpha^k
\end{equation}
In this assumption, $k$ depends on the atmospheric condition and wave
length of the laser. It is empirically estimated to be 0.6$\sim$1.0. 
Substituting this relationship into equation(\ref{eq:lidar}), we obtain
\begin{equation}
\frac{dS}{dR} = \frac{k}{\alpha}\frac{d\alpha}{dR}-2\alpha
\end{equation}
This is a Bernoulli-type equation.

Viezee adopted the boundary condition near the detector and obtained
the following solution\cite{vie73}.
\begin{equation}
\alpha(R) = \frac{X(R)^{1/k}}{\frac{X(R_c)^{1/k}}{\alpha(R_c)}
                - \frac{2}{k}\int^{R}_{R_c} X^{1/k}(r) dr}
\label{eq:viezee}
\end{equation}
Since the second term of the denominator is negative, the denominator
approaches zero with an increase of $R$. This solution
often diverges, and is unstable.

Klett proposed a stable solution by adopting the boundary
condition at the farthest point $R_c$, 
and obtained following solution\cite{klett81}
\begin{equation}
\alpha(R) = \frac{exp[(S(R)-S(R_c))/k]}{\alpha(R_c)+\frac{2}{k}
                \int_{R}^{R_c} exp[(S(R) - S(R_c))/k]dr} \\
        = \frac{X(R)^{1/k}}{\frac{X(R_c)^{1/k}}{\alpha(R_c)}
                + \frac{2}{k}\int^{R_c}_{R} X^{1/k}(r) dr}
\end{equation}
In this solution, the integral term of the denominator becomes larger with
R. Thus the contribution of uncertainty of
the critical value $\alpha(R_c)$ becomes smaller. Therefore this solution
converges on the correct 
value. Using Klett's method, we can analyze $\alpha$, if we can assign
certain values to the parameter '$k$' and the critical value $\alpha(R_c)$.

Furthermore, Fernald proposed a similar solution which considered
atmospheric compositions\cite{fer84}. This solution assumes that $k=1$ and
that scattering 
parameter $S_a\equiv\alpha_M/\beta_M$ is constant. Then $\alpha$ can
be analyzed in successive steps using the Klett method. After
the invention of these methods,
analysis of the LIDAR signal based on elastic scattering
made rapid progress.

\section{Experiment}
\label{Experiment}
%
%
\begin{table}[htbp]
\begin{center}
\begin{tabular}{|l|r|r|r|r|}\hline
\label{tbl:lsr_spec}
Wave length[nm] & 1064 & 532 & 355 & 266 \\ \hline \hline
Pulse Width[nsec] & 6-8 & 5-7 & 4-6 & 4-6 \\ \hline
Energy [mJ] & 50 & 25 & 7 & 5 \\ \hline
Pulse Repetition Rage [Hz] & 10 & 10 & 10 & 10 \\ \hline
Beam Diameter [mm] & 2.75 & 2.5 & 2 & 2 \\ \hline
Beam Divergence [mrad] & 3 & 3 & 2 & 2 \\ \hline
\end{tabular}
\end{center}
\caption{The laser's specification. This laser is a flash-lamp
 pumped, Q-switched, and water cooled Nd:YAG type. At present we use 
 a 355nm wave length.}
\end{table}
%
%
%
%
\begin{figure}[htbp]
\centering
\leavevmode
\epsfxsize=14.5cm
\epsfbox{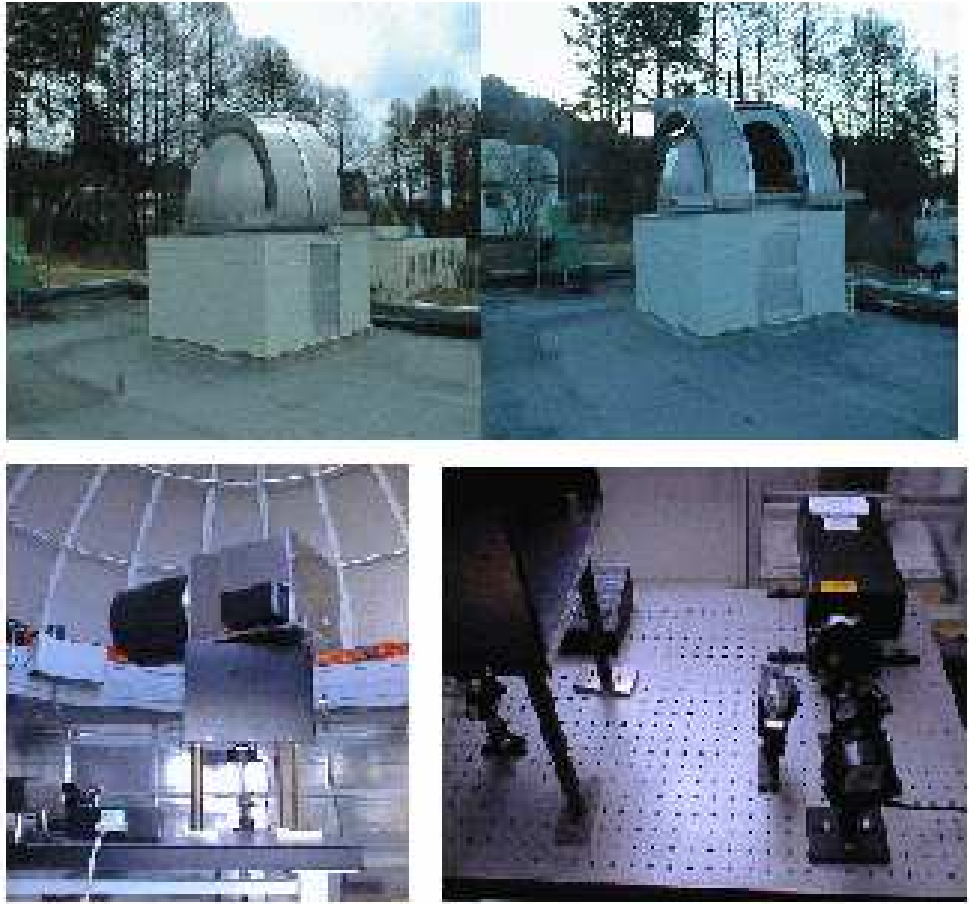}
\caption{Akeno Steerable LIDAR System}
\label{fig:dome}
\end{figure}
%
%
%
%
\begin{figure}[htbp]
\centering
\leavevmode
\epsfxsize=14.5cm
\epsfbox{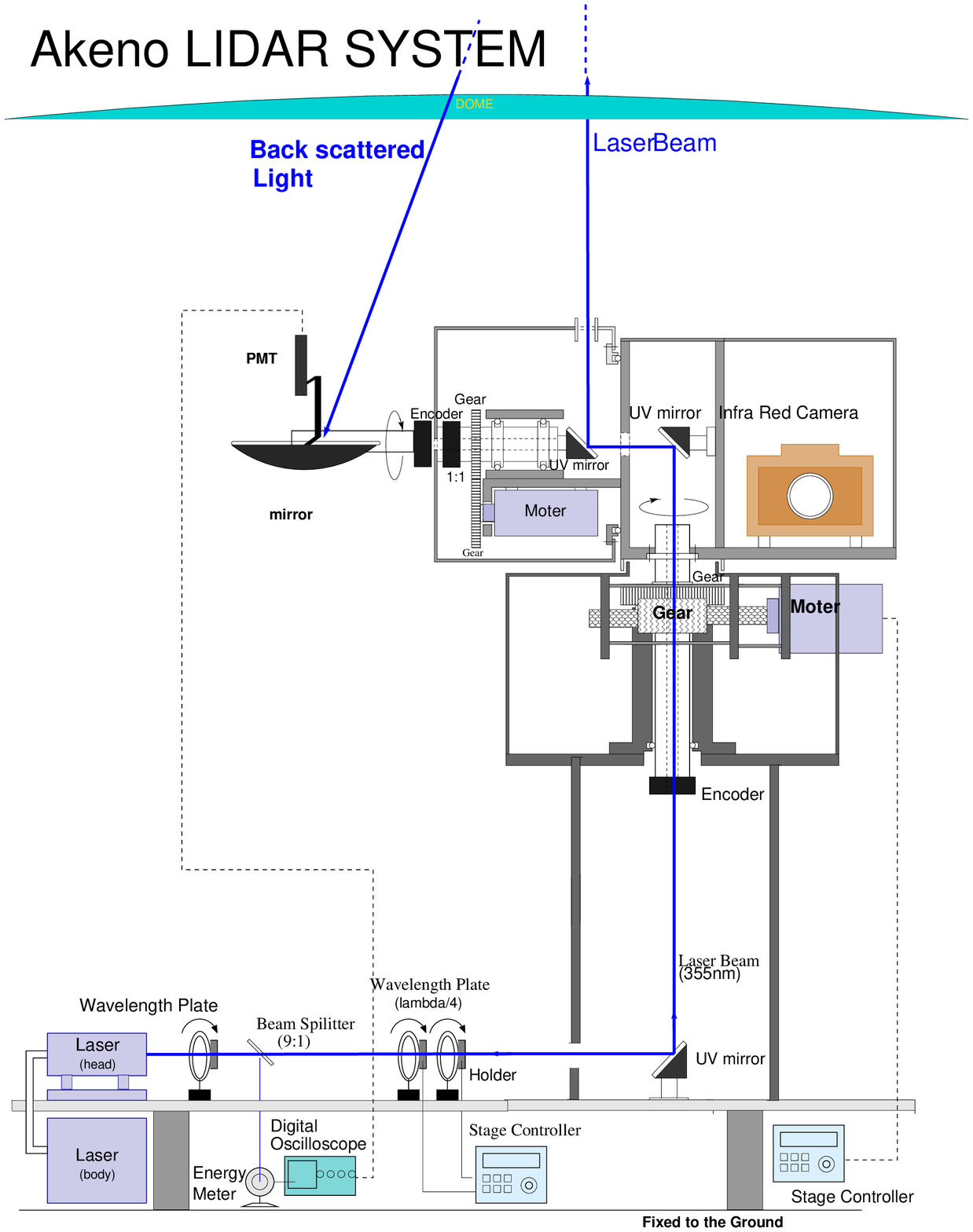}
\caption{Diagram of the steerable LIDAR system developed at Akeno
 observatory.}
\label{fig:altaz}
\end{figure}
%
%

The steerable LIDAR system was developed at Akeno Observatory in
Yamanashi Prefecture, Japan; its geographical coordinate is 900m a.s.l., 
$138.5^\circ$N, $35.78^\circ$E.
This observatory utilizes are many
cosmic ray detectors including AGASA and fluorescence detectors. Using
these detectors, hybrid observation of cosmic rays are performed.

This LIDAR system is situated on the roof of the Akeno Main
laboratory. All of the devices are housed in the astronomical dome
shown in Figure \ref{fig:dome}. This astronomical dome consists of a
cubic-shaped room of 1.8mW$\times$1.8mD$\times$1.0mH and a steerable
dome. The dome window can open to 1m, and its direction must be 
rotated according to the movement of the laser.
Two motors are equipped for opening and closing the dome 
and for dome rotation.
Each motor is controlled by an electronic switch which
operated by a local computer. 

The steerable LIDAR system is illustrated in Figure \ref{fig:altaz}.
Inside the dome, laser and optical devices are installed on a
optical table which is fixed on the concrete pad.
This laser is a flash lamp pumped,
Q-switched, and water cooled Nd:YAG type.
Its specifications are
shown in Table \ref{tbl:lsr_spec}.
The third harmonic beam with a wavelength of 355nm is used in this experiment. 
This wavelength is close to the major lines of air fluorescence light.
The maximum beam intensity is 7 mJ and its maximum repetition rate is 10 Hz. 
The beam intensity, repetition rate, and shooting duration can be 
controlled by the computer through RS232C. 

The laser beam is split into two directions after phase
conversion by the circular polarizer. One has 10\% intensity which is used
for energy calibration, while the other 90\% is transported to the
shooting system. To stabilize the laser and to prevent condensation,
these optical devices including the laser and the optical table
are isolated from the outside atmosphere by a fireproof curtain and
the temperature is kept constant by an air conditioner. 

An alt-azimuth mount is adopted for the shooting system. Encoders are
directly mounted on each axis. The resolution of these encoders is
5/1000 degrees. Each axis is driven by AC stepping motors with 0.0072 degree
steps. The maximum speed of this shooting system is more than 10 degrees/sec. 

An infrared camera is also mounted on the azimuthal axis. Using this
infrared camera, clouds can be recognized as hot regions in the cool night
sky. The typical cloud temperature is higher than a few degrees;
however, the typical night sky temperature is lower than $-10$
degrees. This infrared camera can measure the temperature 
between $-20$ and 20 degrees and its field of view is 20 $\times$ 40 degrees. 
To increase the data rate, the video image from this camera 
is captured as a ppm format image file through a
VIDEO capture board. 

Two mirrors are mounted on the elevation axis. One is a shooting mirror
for the laser beam. The other is a receiving mirror for back-scattered light. 
Maximum beam intensity shot to the sky is about 5 mJ. The
receiving mirror is adjusted to parallel the laser beam. A parabolic
mirror of 16 cm in diameter is adopted for this receiving system. 
Because the shooting and receiving mirrors are mounted on the same part, 
the laser beam and the receiver can be moved simultaneously.
A one-inch PMT is located at the focus of 16cm diameter mirror. 
The field of view is about 1 degree. PMT gain is adjusted to about $10^6$. 
The signal from the PMT is acquired by a 100MHz band-width Digital 
Oscilloscope (TDS33014 made by Tektronix cooperation)
which measures the time profile of the back-scattered light. 
The time range of the signal sweep is set to 10$\mu$sec/DIV and
one sweep consists of 500 wards with a time resolution of 200nsec.
Since the timing of the back-scattered light in the LIDAR system corresponds
to the round-trip time to the target, 
200nsec time resolution is equivalent to 30m in
spatial resolution $\Delta R$. When we set 1 bit to 0.4mV (it corresponds to
10mV/DIV with 256 bits digitizer),
then 1 bit corresponds to about 2 photo-electrons.
An average of 16 shots is calculated by this oscilloscope and
recorded on the local computer. 

All of these systems are controlled by the local computer using a Linux
operating system and operated by the remote computer via a network. 

%
%
\begin{figure}[htbp]
\centering
\leavevmode
\epsfxsize=14.5cm
\epsfbox{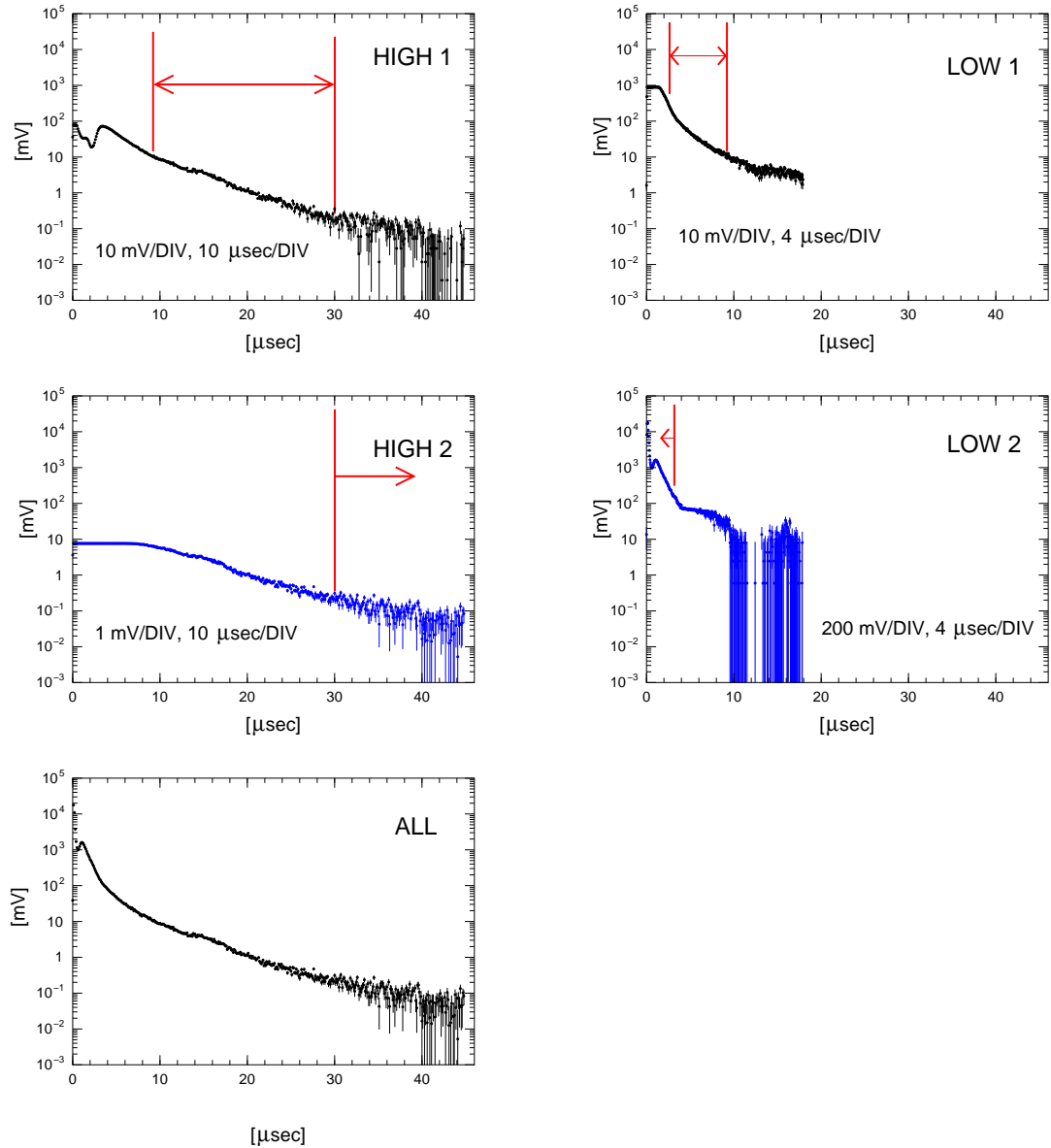}
\caption{Example of a time profile of one direction. An average of 1024
 shots in the vertical direction are plotted. To increase 
 dynamic range, we use two channels: high sensitivity channels (left
 panels) and low sensitivity channels (right panels). Laser intensity
 is 5mJ for the high sensitivity channel and 0.4mJ for the low sensitivity
 channel. The amplitude of the low sensitivity channel is normalized by
 laser intensity.  Each channel is measured in two 
 oscilloscope ranges  as indicated in each panel. The regions 
 indicated by dashed lines are used for the analysis. The lower panel indicates
 the average of the four channels.}
\label{fig:akeno010423_1_bs}
\end{figure}
%
%
%
%
\begin{figure}[htbp]
\centering
\leavevmode
\epsfxsize=14.5cm
\epsfbox{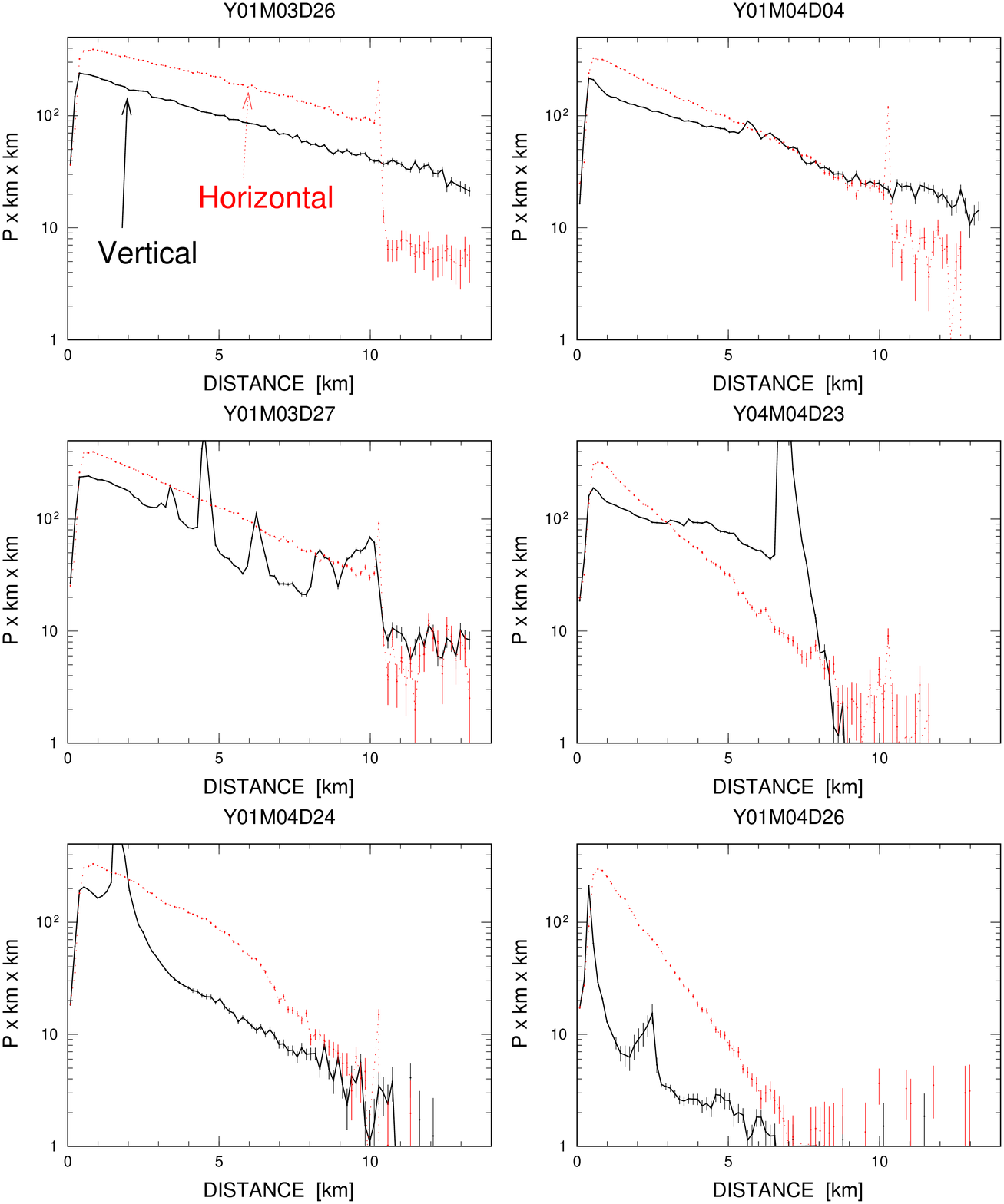}
\caption{X-axis is the distance from the laser. Y-axis is the pulse height 
 multiplied by the square of the distance ($X(R)$). Dotted lines indicate 
 the horizontal measurement. Solid lines indicate vertical measurement. 
 The upper left panel shows a result obtained on a clear night. 
 The lower right panel shows a result obtained on a cloudy night. 
}
\label{fig:bs2}
\end{figure}
%
%

\section{Observation}
\label{Observation}

The Telescope Array is planed to build 10 stations with 40km separation in
the Utah desert USA. The location of those stations will be spread over 
a very wide
area, their operation should be done remotely via a network
linked by microwave or optic fiber. This would constitute a severe constraint
for our atmospheric monitoring system. For this reason, we developed a fully
automated system at Akeno observatory.
For example, if it begins to rain during the observation, the system is
shut down automatically.

The intensity of the back-scattered light reaching the detector 
from a scattering point 
is proportional to the inverse of the square of the distance, the density of
the atmosphere, and the intensity of the laser beam. For this reason, 
the amount of returned light from near the detector is huge 
and that from a distant point is faint.
Therefore a very wide dynamic range, typically 4 orders of magnitude, 
is required for the light receiving detector. 
If we use a high
intensity laser beam, the PMT is saturated by photons scattered
near the detector. If it is weak, the number of measurements have to be
increased in order to measure the signal from a distant place,
thus lengthening the observation time.

To increase the dynamic range, we use  low and high
sensitivity channels. Each channel uses two settings on the digital
oscilloscope as illustrated in Figure \ref{fig:akeno010423_1_bs}.
For the high sensitivity channel, the laser beam is shot at 5
mJ into the sky and oscilloscope ranges are adjusted to 1mV/DIV and
10mV/DIV. In this case, the PMT is saturated by scattered photons  
near the detector and it takes more than 9 $\mu$sec to recover
sensitivity. Data points with signal amplitudes of less than 2 mV
are not used in the analysis in order to obtain a reasonable signal 
to noise ratio, and to reduce the uncertainty of the pedestal. 
This value corresponds to about 1
p.e., i.e., 5 bits for 1mV/DIV.
For the low sensitivity channel, the laser intensity is reduced
to about 1/12 and the oscilloscope ranges are adjusted to 10mV/DIV
and 200mV/DIV.  
In this case, the PMT recovers by 8 $\mu$sec. Using this
channel, we can measure the back-scattered light from a distance between 
1.2 km and 3 km. 

The signal profiles of 16 shots are averaged by the digital
oscilloscope and then recorded on the local computer. 
This measurement is repeated 48  
times in vertical and horizontal directions and 16 times for every 5
degrees between 85 and 5 degrees.
In other words, 1024 shots are measured in the
vertical and horizontal directions, and 256 shots are measured 
in other directions. 

Examples of measurements are shown in Figure \ref{fig:bs2}. 
In this figure, the results in the vertical and horizontal directions 
are shown.
In the horizontal measurement, if the atmosphere depends on only altitude (
meaning that the atmosphere has a 1-D structure and is uniform), 
$S(R)\equiv log(X(R))$ is in proportion to
R and extinction coefficient $\alpha$ is analyzed in terms of the slope
using equation(\ref{eq:slobe}). If
the atmosphere is not uniform in the horizontal direction, 
the slope is not constant, depending on the atmospheric conditions. 
We can see S(R) decline linearly to 10 km under good weather 
conditions as shown in Figure.
There is a mountain at a distance of 10km and the light scattered by
this mountain can be seen in this figure. The atmosphere is not uniform
under bad weather and S(R) does not decline linearly. 
In measurements in the vertical direction, it can be clearly seen
that clouds cause the magnitude of the signal strength to vary rapidly.
We can measure the atmosphere up to an altitude of 12km under a clear sky.

\section{Analysis}
\label{Analysis}
%
%
\begin{figure}[htbp]
\centering
\leavevmode
\epsfxsize=14.5cm
\epsfbox{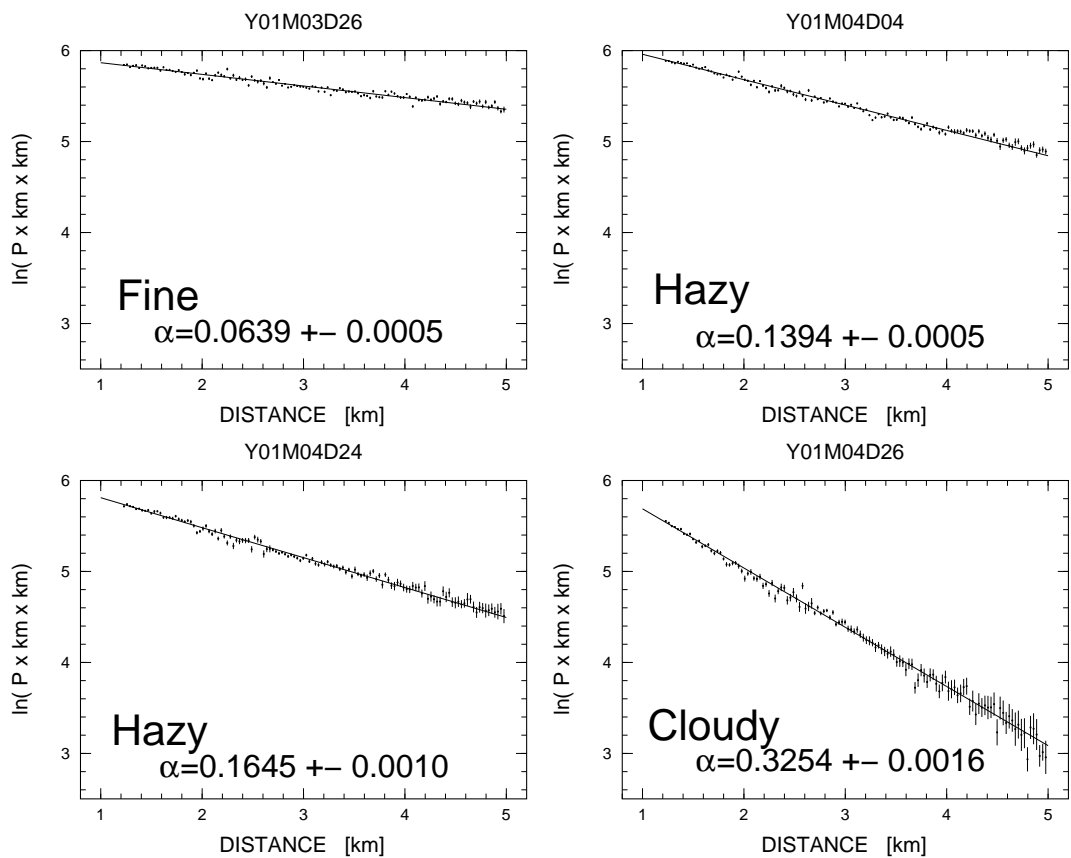}
\caption{Example of horizontal measurement. Regions from 1.2 km to 5 km
 are plotted. The upper left panel represents the clearest night, 
 and the lower right panel represents a cloudy night. 
 Solid line is result of fitting by straight
 line. The extinction coefficient $\alpha$ can be estimated using the slope
 of this line. The estimated $\alpha$ is indicated in each panel.}
\label{fig:bs4}
\end{figure}
%
%
%
%
\begin{figure}[htbp]
\centering
\leavevmode
\epsfxsize=14.5cm
\epsfbox{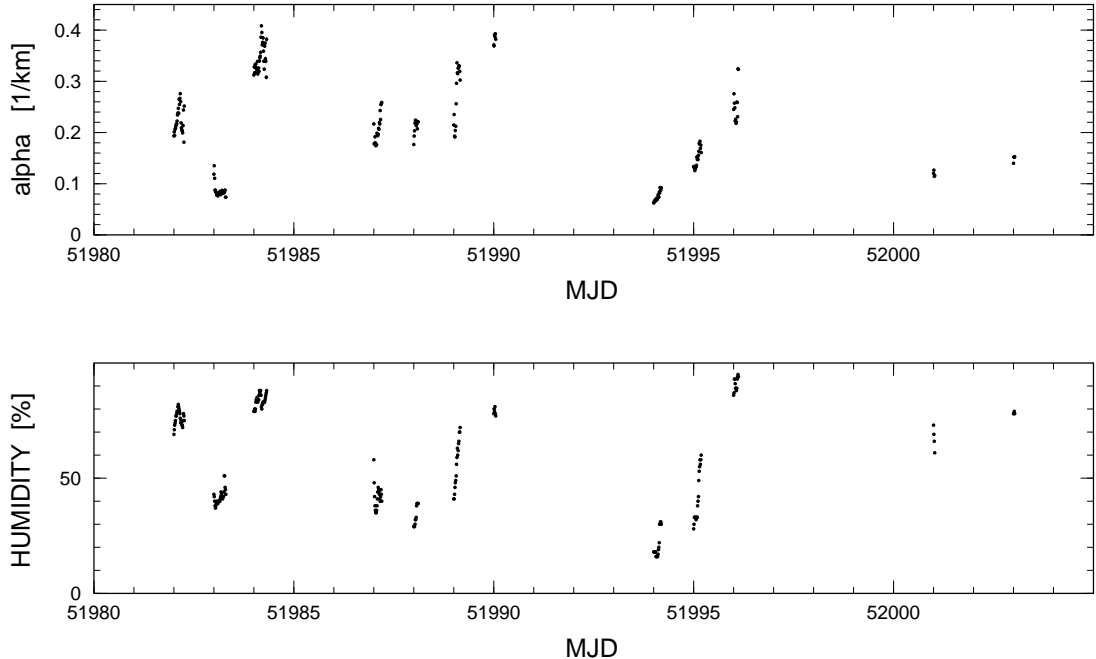}
\caption{Upper panel shows estimated $\alpha$ at the level of the
 detector as a function of observation time. Lower panel shows
 humidity measured simultaneously.}
\label{fig:al_hor}
\end{figure}
%
%
%
%
\begin{figure}[htbp]
\centering
\leavevmode
\epsfxsize=14.5cm
\epsfbox{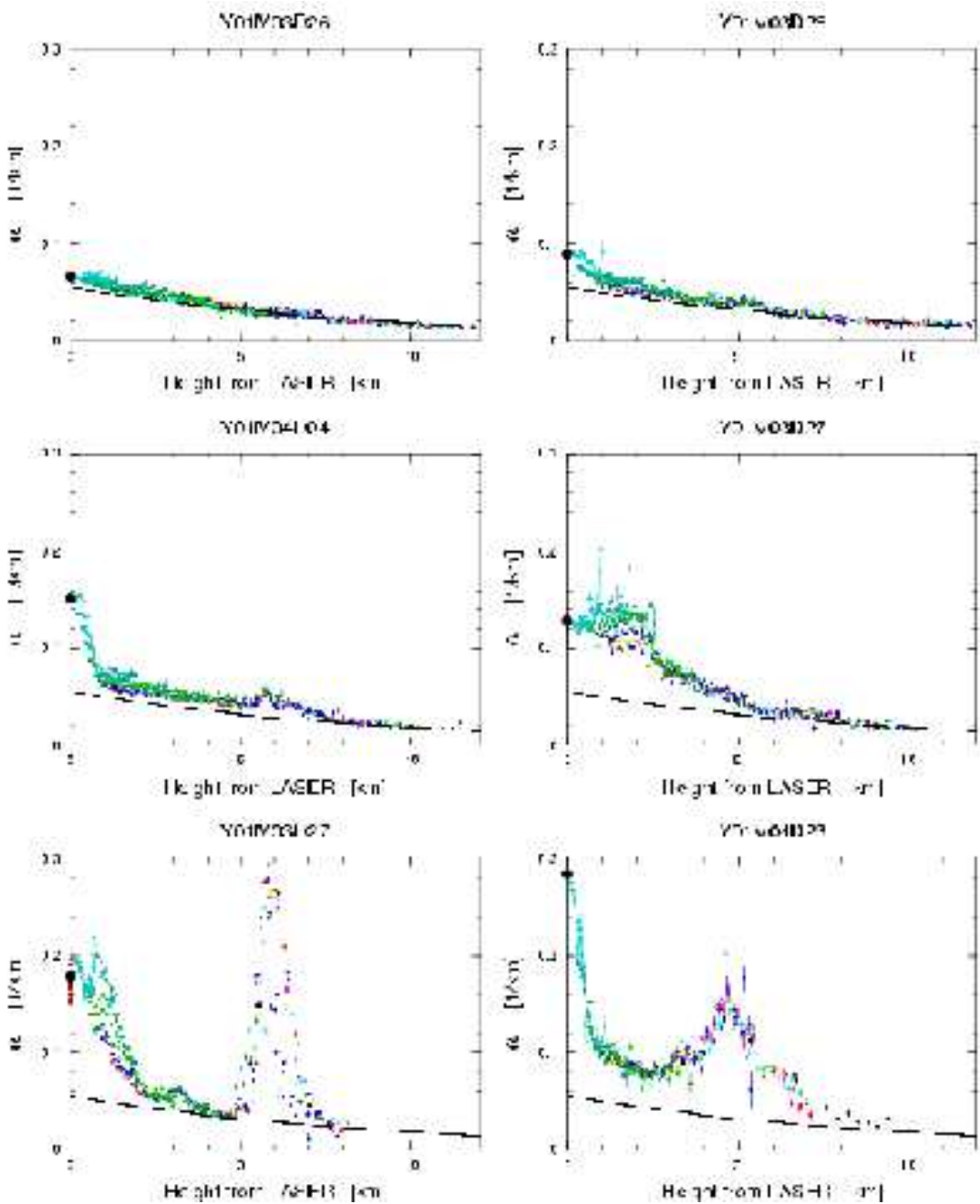}
\caption{Estimated $\alpha$ using Klett's method as a function of
 altitude. Different colors correspond to different zenith 
 angles. Solid lines indicate the expected value from Rayleigh
 scattering. Black dots at height 0 indicate the measured value in
 horizontal shot. The upper left panel shows the results of the clearest night.
 The lower left panel shows a cloud at around 6km altitude. This cloud
 obscures the signal from a higher altitude. Because the boundary condition 
 cannot be determined, the absolute value cannot be analyzed correctly. 
}
\label{fig:bs9}
\end{figure}
%
%
%
%
\begin{figure}[htbp]
\centering
\leavevmode
\epsfxsize=14.5cm
\epsfbox{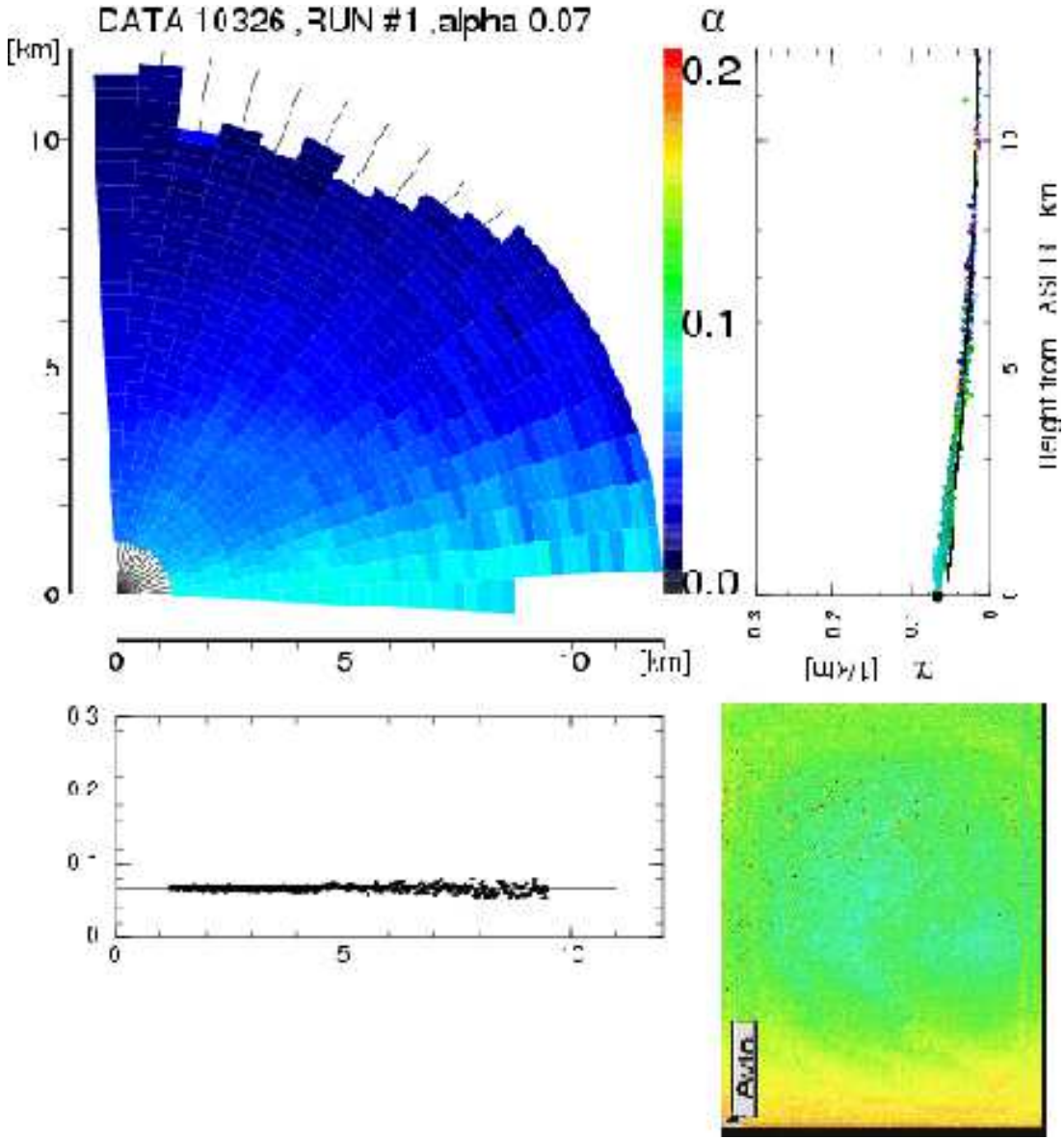}
\caption{An example of estimated extinction coefficient $\alpha$. The upper 
 left panel shows a two-dimensional map of $\alpha$ inside a 12 km
 radius circle with LIDAR in its center. Estimated $\alpha$ in horizontal
 measurement is indicated above this panel. The upper right panel is the same 
 as that in the previous figure. The lower left panel shows the
 estimated extinction coefficient in horizontal shots. The solid
 line indicates the result of fitting. The lower right
 panel shows an image taken by the infrared camera. This data regards the 
 clearest night.}
\label{fig:akeno010326_1_bs10p1}
\end{figure}
%
%
%
%
\begin{figure}[htbp]
\centering
\leavevmode
\epsfxsize=14.5cm
\epsfbox{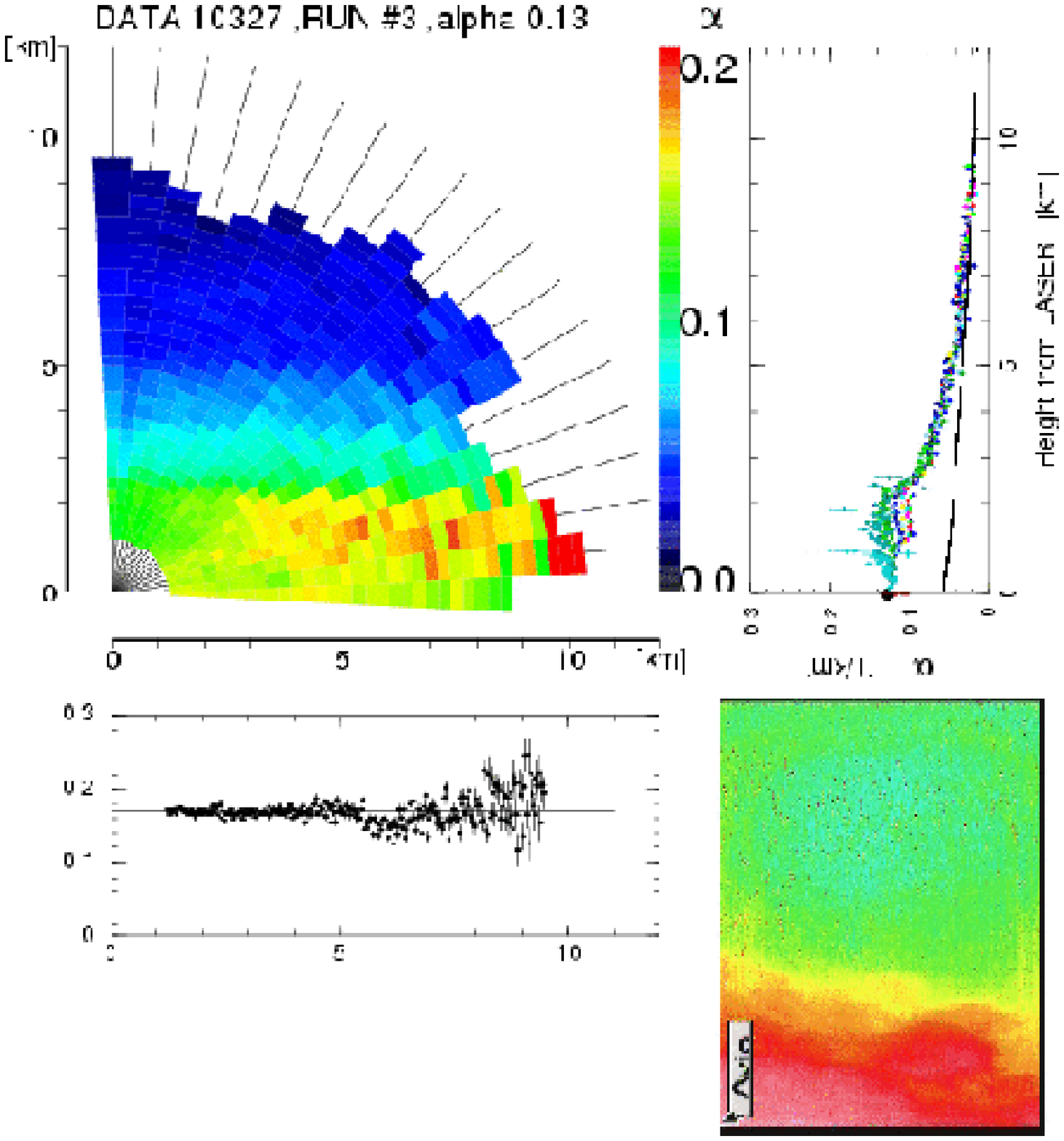}
\caption{Data of the next night. 
 The expression is the same as in the previous 
 figure. The atmosphere was hazy at low altitudes.}
\label{fig:akeno010327_1_bs10p1}
\end{figure}
%
%

In this section, we discuss how to estimate the transmittance of the
atmosphere using the results of the measurements described above. 

One of the simplest cases is horizontal measurement. If the atmosphere is 
uniform, $\alpha$ can be analyzed using equation(\ref{eq:slobe}) at the level
of the detector as mentioned previously. 
Figure \ref{fig:bs4} shows examples of 
this horizontal measurement. Data which is comparatively near the
detector is used. It can be seen that the
attenuation of the laser beam clearly depends on the weather. It can
also be confirmed 
that the atmosphere is uniform in this region. Figure \ref{fig:al_hor} shows 
the time profile of $\alpha$ in comparison with humidity which is measured
simultaneously. Of course, the transmittance of the atmosphere depends not
only on humidity; however, in this figure we can see that variation 
of $\alpha$ follows 
tiny variations of humidity. This method appears to be
very convincing and accurate. 

As described section \ref{ss:lidar_eq}, the LIDAR equation cannot be
solved because it  
has two variable parameters. A solution which uses an assumption of
the empirical relationship between $\alpha$ and $\beta$ is proposed by
Klett. To estimate $\alpha$ using Klett's method, we need to determine
two parameters: one being the critical value at the highest altitude point, the
other being parameter $k$ which represents the relationship between $\alpha$
and $\beta$. To determine these parameters, we use a simple assumptions as
follows:
\begin{itemize}
 \item In a measurement in the vertical direction, the smallest
       $\alpha$ corresponds to the value expected from Rayleigh
       scattering. If there are no clouds, this value can be estimated at
       its highest point.
 \item In a measurement in a different direction, a smaller $\alpha$
       (meaning the value in a cloudless region) corresponds to the
       $\alpha$ which is measured at a direction of +5 degrees at
       the same altitude.
 \item $\alpha$ at points lower than 100m can be estimated by the
       $\alpha$ obtained in the horizontal measurement.
\end{itemize}
      In the first assumption, we estimate the critical value based on
        Rayleigh scattering. This means that we assume
        the atmosphere is clear and the effect of Mie scattering can be 
	ignored at high altitudes, which is about 10 km with the present 
	system.
        Transmittance is analyzed in successive steps from the critical point
        to the lower point. For this reason, errors in the critical value 
	affect the measurement of the lower points. 
        This assumption is reasonable during good weather conditions. 
	However, if the weather is not clear, this assumption causes
	significant systematic error.

        At small elevation angles, it became difficult to measure the 
        transmittance at high altitude. 
        In the second assumption, the measurement at larger 
        elevation angle is used for estimation of the critical value of the 
        smaller one. In other words, we assume uniform atmosphere at the 
	same altitude for the measurements of adjacent elevations.
	If atmospheric conditions changes rapidly or are localized, it
        is impossible to estimate the alpha because of this assumption.

	In the Telescope Array experiment, we expect that no observations
	will be performed or that the data will be discarded under cloudy
	conditions or if the atmospheric conditions in the field of view
	change rapidly.	Under these conditions we estimate 
        the observation's duty cycle to be 7$\sim$10\% 
	at the  Utah site. For this reason, the assumptions made in this 
	analysis should be acceptable.
Based on these assumptions, we can determine the parameter '$k$' and the
critical value.

Figure \ref{fig:bs9} shows the examples of estimated $\alpha$ as a function
of altitude at each zenith angle. As we can see, the atmosphere is almost
uniform if there are no clouds. In the lower left panel, a cloud can be seen
at around an altitude of 6 km. This cloud obscures the signal from higher
altitudes. Since the boundary condition at high altitudes cannot be determined,
the absolute value of $\alpha$ cannot be analyzed correctly.
Figure \ref{fig:akeno010326_1_bs10p1} shows the results obtained on a very 
clear night. The atmosphere of this night can be almost 
explained by pure molecular at high altitude. Also the atmosphere  was 
very uniform as is illustrated. 
It can be clearly seen that the statistical error is very small at all
points. For example, the transmittance between the detector and a 10 km
distance in the vertical direction is estimated to be 73.27\% with
a statistical error of 0.05\%.
Figure \ref{fig:akeno010327_1_bs10p1} shows the result of measurements 
taken on the next night. The weather was hazy and the atmosphere was not
uniform at lower heights. The transmittance between the detector and 10km
in the vertical direction is 57.91\% with a statistical error of
0.07\% It takes approximately 20 minutes 
to measure one azimuthal direction.

\section{Systematic Error}
\label{Systematic Error}
%
%
\begin{figure}[htbp]
\centering
\leavevmode
\epsfxsize=10cm
\epsfbox{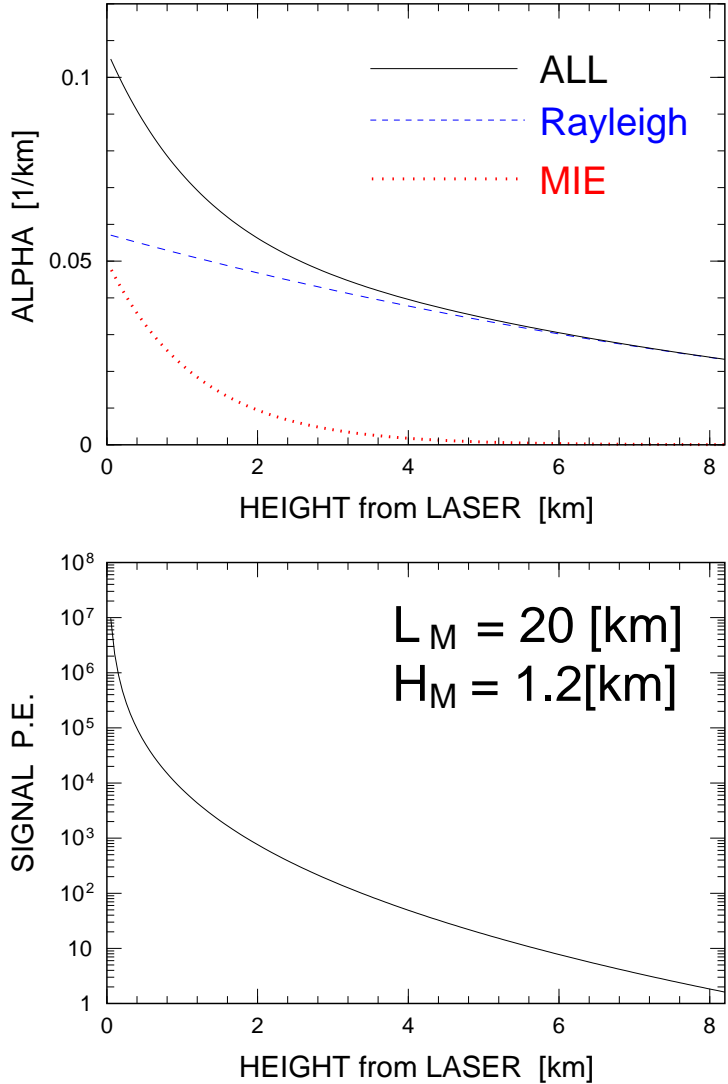}
\caption{An example of the atmospheric model used for simulation.
 Typical Mie parameter values are used for this simulation as
 indicated in the figure. The diameter of the receiver mirror is 
 1.5m, laser intensity is 10mJ, and the laser elevation angle is 10 degrees.
 In the upper panel, dashed and dotted lines indicate the expected $\alpha$ 
 from Rayleigh and Mie scattering, respectively. The solid line indicates 
 the expected $\alpha$ from both types of scattering.
 The lower panel shows the expected number of
 photo-electrons detected using LIDAR.}
\label{fig:lidar_sim3}
\end{figure}
%
%
%
%
\begin{figure}[htbp]
\centering
\leavevmode
\epsfxsize=14.5cm
\epsfbox{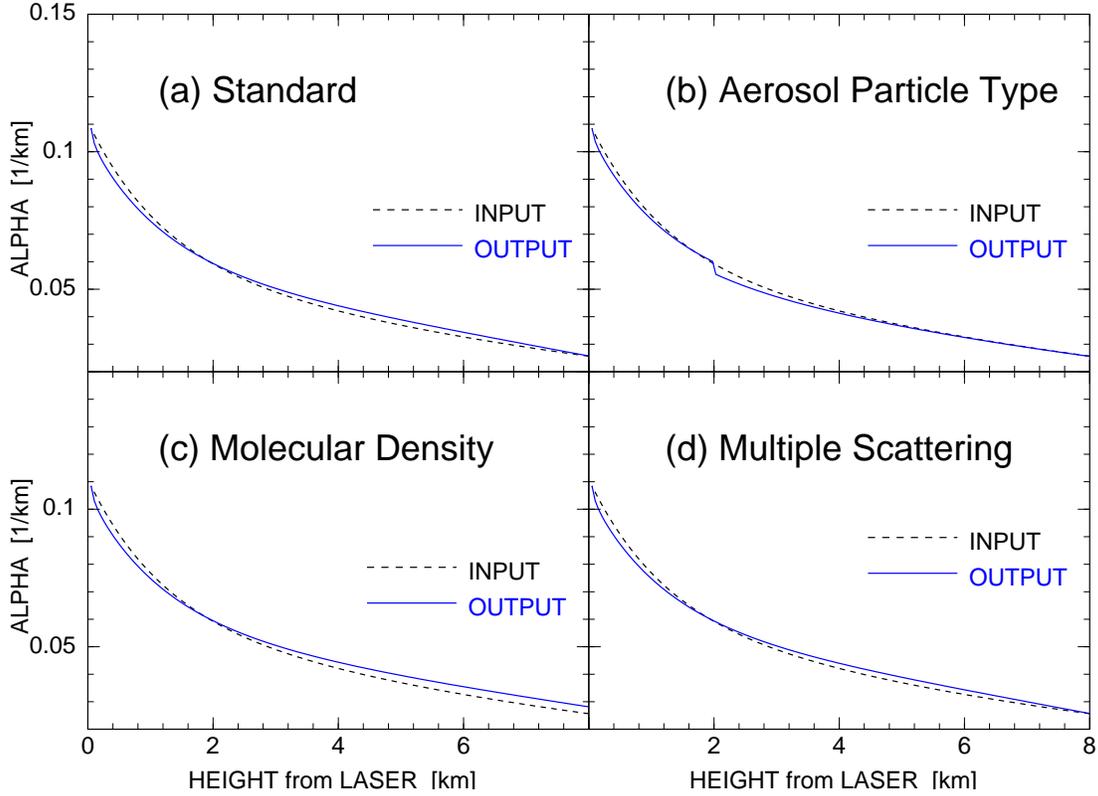}
\caption{Examples of results of the simulation. 
        Dashed lines indicate the true value
        of alpha which is calculated from each atmospheric model.
        Solid lines indicate the estimated alpha based on Klett's method 
        under the assumption of that alpha is known at the level 
        of the detector.
        (a) The standard atmospheric model is used as in the previous figure.
        (b) $\beta_M$ is changed from 0.1 to 0.05 at 2km height. In this case,
        the total cross section which proportional in $\alpha$ is the same 
	as the standard
        model but the phase function of Mie scattering changes rapidly.
        (c) Molecular density is overestimated at its highest point.
        (d) Multiple scattering is considered based on a simple assumption.
        In this case, 10\% of the observed light is assumed to be
        multiple scattered light. The details of each parameter is in the
        text.
	}
\label{fig:lidar_sim5}
\end{figure}
%
%
%
%
\begin{figure}[htbp]
\centering
\leavevmode
\epsfxsize=14.5cm
\epsfbox{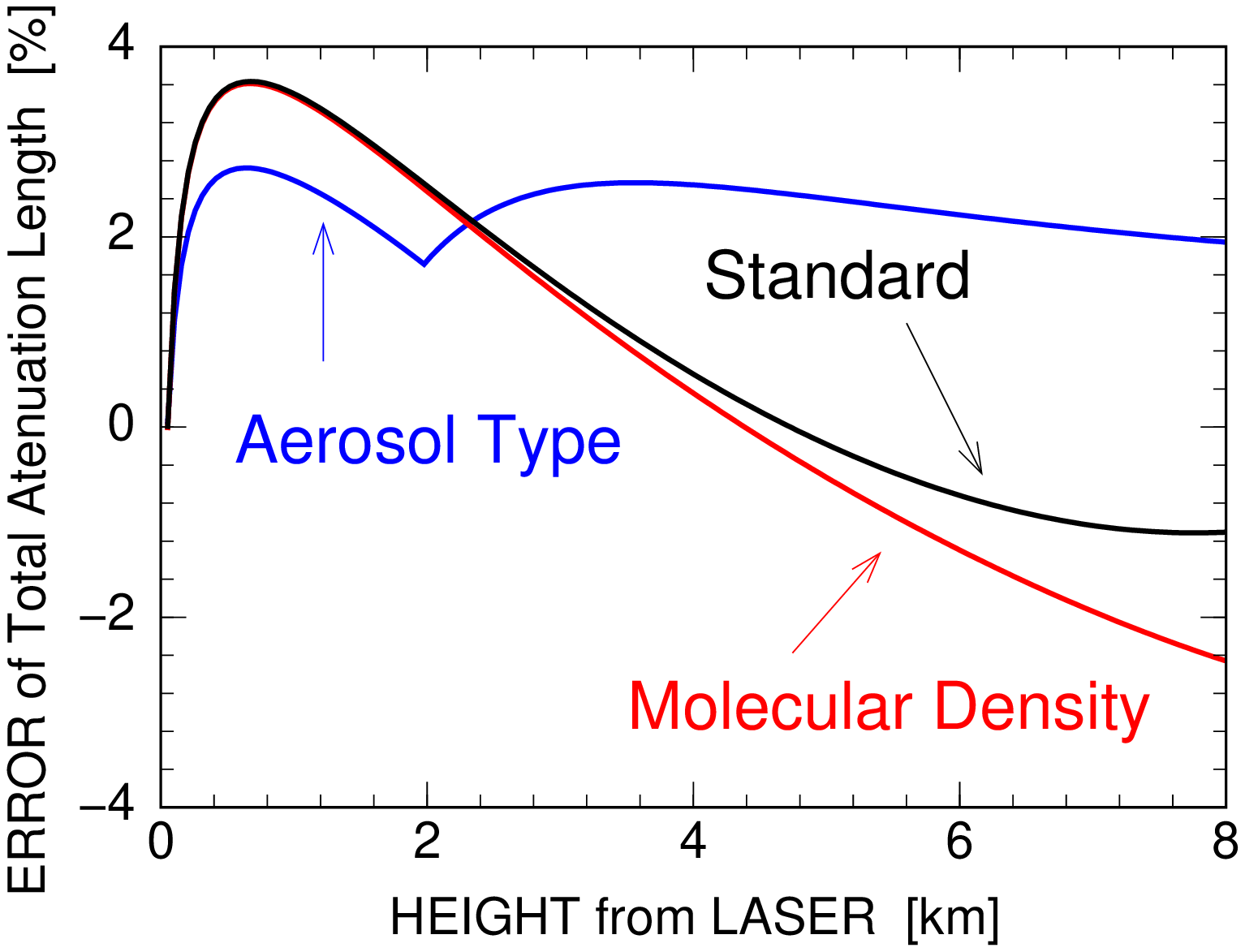}
\caption{The systematic error of total attenuation length between the laser
        and the measurement point as a function of height. The input 
	atmospheric models are the same as in the previous figure. }
\label{fig:lidar_sim6}
\end{figure}
%
%

	The systematic error in the present analysis based on the Klett's 
	method has been investigated using a simple Monte-Carlo simulation
	\cite{mat01}\cite{mat02}.
	In this simulation, the signal profile of back-scattered light 
	is calculated using a simple atmospheric model.
	The number of Rayleigh-scattered photons follows molecular density
	only, while Mie scattering is calculated by assuming
	Mie parameters: scale height $H_M$, horizontal attenuation length
	$L_M$, and back-scatter coefficient $\beta_M$.
	An example of the atmospheric model with typical Mie parameters
	($L_M$=20km, $H_M$=1.2km, $\beta_M$=0.05) is shown in Figure 
	\ref{fig:lidar_sim3}.
	In this calculation, the laser's elevation angle is assumed to be
	10 degrees. 

	This artificial data is analyzed using the same programs as above,
	employing real data analysis. In Figure \ref{fig:lidar_sim5}, 
	examples of the results are shown.
	The expected alpha which is calculated based 
	on the input atmospheric model and the estimated one using 
	Klett's method are compared. 
	In Figure \ref{fig:lidar_sim5}.a the standard atmospheric model is
	assumed with typical Mie parameters as in the previous figure.

	In Figure \ref{fig:lidar_sim5}.b, Mie parameters are the same as the 
	standard 
	model except for $\beta_M$ which is transformed from 0.1 to 0.05
	at 2 km height. The Klett's assumption of a power-law relationship 
	between	$\alpha$ and $\beta$ does not take into account 
	variation of the aerosol particle types. If the aerosol particle size 
	is large, forward-scattered light is predominated in Mie scattering.
	If it is small, the phase function of Mie scattering becomes similar 
	to that of Rayleigh scattering.
	In this panel the total cross section is the same as the  standard 
	model, but the phase function of the Mie scattering changes rapidly 
	at 2 km height. The changing of aerosol types can cause significant 
	systematic errors in this analysis as we can see.

	In Figure \ref{fig:lidar_sim5}.c, all Mie parameters are the
	same as those of the standard model. But the critical value which 
	is determined based on Rayleigh scattering at the furthest point is 
	assumed to have a 10\% systematic error.
	This critical value is estimated under the assumption that all of the
	backscattering at the highest point is Rayleigh scattering.
	In this case, this critical value can be estimated based on molecular
	density which depends on temperature, pressure and the type of 
	atmospheric model used.
	In this panel, we assume that the critical value is overestimated
	beyond the true value by 10\%. A similar situation can arise when
	we ignore the effect of Mie scattering to obtain the critical value 
	under an unclear atmosphere.

	In Figure \ref{fig:lidar_sim5}.d, all Mie parameters 
	are the same as in the standard model.
	But multiple scattered light is considered under the assumption 
	that 10\% 
	of the observed backscattering light is multiple scattered light.
	A more complete study of the effect of multiple scattering lies 
	outside the scope of this paper. 
	There is no effect to the estimated value under this simple model.

	Figure \ref{fig:lidar_sim6} shows the systematic error 
	of the total attenuation length
	between the laser and measurement point as a function of height.
	According to these figures, the systematic error caused by Klett's
	method is less than 5\% at all of the measurement points in
	the standard atmospheric mode. 
	Most significant effect is caused by variation of the aerosol type.
	These systematic errors should be dependent on the laser's elevation 
	angle, the Mie parameters, and the type of atmospheric model used.
	Further investigation of the simulation study is described
	in \cite{mat02}\cite{mat03}.

The detector constants (for example: mirror reflectivity, PMT gain, beam
intensity, and so on) do not contribute to the systematic errors in this
analysis except for the linearity of the PMT. The effect of multiple
scattering is ignored in this measurement. The most significant
systematic error is caused by the uncertainty of the critical value. The 
altitude of 10 km, where we established the boundary condition, 
may not be high enough to 
ignore the effect of Mie scattering. This problem can be solved if we
use a larger mirror or a higher intensity laser.

\section{Summary}
\label{Summary}

We have developed a steerable LIDAR system for atmospheric monitoring
by the Telescope Array. The system consists of a 5 mJ pulse laser and
16 cm diameter mirror. Using this system, a technique for atmospheric
monitoring was developed. 

First, the extinction coefficient $\alpha$ at the level of the detector 
is measured. Then $\alpha$ is estimated at all directions using Klett's
method. The transmittance of the night sky to a distance of more than 10
km can be measured successfully. The statistical error in this analysis
is less than 2\% under a clear sky. It takes approximately 20 minutes 
to measure one azimuthal direction.

The systematic error which is caused by this method of analysis is
estimated to be approximately  0.5\% in a typical atmospheric model. 
The detector 
constants do not contribute to the systematic errors in this analysis
except for the linearity of the PMT. The most significant systematic
error is caused by uncertainty of the critical value and variation
of the aerosol type. Some of these problems can
be solved if we use a larger mirror or a higher intensity laser.

Based on these observational results, we are considering future plans
regarding atmospheric monitoring for the Telescope Array. The signal to noise 
ratio is in proportion to beam intensity, mirror diameter, square root
of observation time, and transmittance.
$5mJ \times 16 cm \times \sqrt{20 min} = 358[mJ \cdot cm \cdot min^{1/2}]$ 
is required to measure
to a distance of 10km under a clear sky (in this case, the transmittance
is about 60\% in the vertical direction). To measure more than 50 km, we need
$358 \times 5^2 / 0.6^5 = 115000 [mJ \cdot cm \cdot min^{1/2}]$. 
This means that a 270 mJ laser is
necessary for a 3 m diameter mirror if we want to measure one azimuthal
direction in 2 minutes. This value can be reduced by a narrow band filter,
small field of view, and high altitude location
since the main component of the noise is night sky background light.
For a telescope with a 1 degree field of view and F=1,
the observed night sky background is about 5 MHz in the ultraviolet region.

	For the operation of the Telescope Array detector, 
        independent and redundant measurements of the atmosphere should be 
        done for the sake of verification and cross-checking.
        In fact, several measurements being performed by the Telescope Array and
        the HiRes groups
	\cite{he1.71}\cite{he1.712}\cite{he1.72}\cite{he1.73}\cite{he1.74}\cite{he1.75}\cite{he1.76}. 
        For example, These include the measurement of the large angle 
	scattered light
        of a polarized laser beam and radio-controlled xenon flashers using 
        the HiRes detector, 
        monitoring the transmittance of the air around the detector 
        using the Nepherometer, measurement of the zenith angle dependence of 
        star light and so on.
        By combining these measurements and the LIDAR method, more reliable 
	measurements can be performed in the future.

\begin{ack}

This work is supported by grants in aid \#12304012 and \#11691117 
for scientific research from the JSPS(Japan Society for the Promotion of
Science). The authors thank Dr. Lawrence R. Wiencke, Dr. Michael
E. Roberts, and Prof. John A.J.Matthews for their thoughtful discussions.

\end{ack}



\begin{thebibliography}{99}


\bibitem{law99} L.R.Wiencke, {\em et all} : NIM A {\bf 428}(1999)593-607
\bibitem{og1} M.Chikawa, {\em et all} : Proceedings of the 26th ICRC (1999)OG4.5.17
\bibitem{og2} R.Gray, {\em et all} : Proceedings of the 26th ICRC (1999)OG4.5.04
\bibitem{og3} N.Hayashida, {\em et all} : Proceedings of the 26th ICRC (1999)OG4.5.05
\bibitem{og4} J.R.Mumford, {\em et all} : Proceedings of the 26th ICRC (1999)OG4.5.10
\bibitem{vie73}
	W.Viezee, J.Oblanas, R.T.H.Collis:SRI report AFCRL-TR-73-0708(1973)
\bibitem{klett81} J.D.Klett : Appl.Optics,{\bf 20}(1981)211-220
\bibitem{fer84} F.G.Fernald : Appl.Optics,{\bf 23}(1984)652-653
\bibitem{mat01} John A.J.Matthews : private communication(2001)
\bibitem{mat02} John A.J.Matthews : HiRes/Pierre Auger Note GAP-2001-046
\bibitem{mat03} John A.J.Matthews : HiRes/Pierre Auger Note GAP-2001-051
\bibitem{he1.71} M.D.Roberts, {\em et all} : Proceedings of the 27th ICRC (2001)HE1.07
\bibitem{he1.712} M.D.Roberts, {\em et all} : Proceedings of the 27th ICRC (2001)HE139
\bibitem{he1.72} M.Sasano, {\em et all} : Proceedings of the 27th ICRC (2001)HE1.07
\bibitem{he1.73} T.Yamamoto, {\em et all} : Proceedings of the 27th ICRC (2001)HE1.07
\bibitem{he1.74} L.R.Wiencke, {\em et all} : Proceedings of the 27th ICRC (2001)HE140
\bibitem{he1.75} M.Chikawa, {\em et all} : Proceedings of the 27th ICRC (2001)HE144
\bibitem{he1.76} R.W.Clay, {\em et all} : Proceedings of the 27th ICRC (2001)HE143


\end{thebibliography}
\end{document}